\documentclass{article}

\begin{document}
\newcommand {\thetap}{$\Theta^+$ }

\title{Experimental Outlook for the Pentaquark\\
}

\author{Ken Hicks \footnote{email: hicks@ohio.edu} \\ 
{\it Department of Physics and Astronomy, Ohio University,} \\
{\it Athens, Ohio 45701, USA }
}

\maketitle


\begin{abstract}
A critical look is taken at both positive and null evidence 
for the \thetap pentaquark.  Potential problems with experiments 
will be discussed and the question of what conclusion can be 
drawn from both the positive and the null results 
is examined.  First the question of existence of the 
\thetap pentaquark is considered, followed by a discussion 
of new experiments that are either planned or in progress 
to answer questions about its mass, width and isospin. 
Finally, indirect evidence for the parity of the \thetap 
is examined, and suggestions for experiments to measure 
its parity directly are given.

\end{abstract}

\section{ Introduction }

The possible existence of the \thetap remains one of the 
most exciting topics in nuclear physics.  To date there 
are over ten experiments\cite{thetaexpt} 
with evidence for this state, and a similar number of 
high-energy experiments\cite{nullexpt} that see no evidence 
for the \thetap even though other states, such as the 
$\Lambda(1520)$ hyperon resonance, are seen clearly.  
Furthermore, the \thetap is not seen in $e^+ e^-$ 
collisions\cite{eeexpt} where, for example, the 
J/$\psi$ or the $\psi(2s)$ could decay to 
a \thetap and anti-\thetap final state.  On the other 
hand, in the latter two cases we can only guess at the 
\thetap production mechanism (or lack thereof).
In contrast, several experiments at medium energies 
have exclusive final states and production cross sections 
that can be calculated\cite{thetacalc} within a theoretical 
model, given some reasonable assumptions.

Can the contrasting results of these experiments be understood?  
This is the central theme of the present paper, along with an 
overview of experimental results expected in the near future.  
If the \thetap exists, it is important to design experiments that 
can accurately measure its width, spin and parity.  First, 
a critical look is taken at both positive and null experimental 
results for the \thetap pentaquark. 

\section{ Does the \thetap Exist? }

Virtually all experiments are subject to some criticism.  
It is quite difficult to understand the 
systematic uncertainties in a measurement, and this is 
especially true when the statistics are limited.  The 
experiments with evidence for the \thetap have low 
statistics, and the background under the peaks may not 
be completely understood.  As a result, the statistical 
significance of the evidence has been questioned. 
It makes sense to focus on the most reliable experiments 
to answer the question of whether the \thetap exists.

The first experiments from the LEPS\cite{leps}, DIANA 
\cite{diana}, CLAS\cite{stepanyan} and SAPHIR\cite{saphir} 
collaborations were ground-breaking, but each has 
some weakness.  The LEPS experiment had only 19 counts 
in the peak on top of a background that was 17 counts, 
so detailed studies of the systematics of the background 
and the Fermi motion correction were difficult.  (New 
data from LEPS with more statistics will be presented 
below.)  The DIANA experiment is hampered by background 
from kaon charge-exchange reactions, and not enough 
detail is given in their paper to show how the cuts 
they employ to reduce this background affect the mass 
spectrum with the \thetap peak, which is concentrated 
into a single bin.  The CLAS data was the first exclusive 
reaction on the \thetap but requires a complicated 
mechanism with secondary-scattering to give energy to 
the proton, which would otherwise be a spectator. As a 
result, the shape of the background under the \thetap 
peak is difficult to estimate and may include kinematic 
reflections\cite{dzierba}. The SAPHIR collaboration was 
the first to publish for the $\gamma p \to K_s^0 K^+ n$ 
reaction, but the large cross section they estimated from 
their measurement conflicted with data for the same reaction 
from CLAS\cite{hadron03}. A re-analysis of the SAPHIR 
data\cite{ostrick} suggests a smaller cross section but 
is still under study.

Following the first reports, several experiments measured 
the invariant mass of the $K_s^0$ and a proton, which showed 
a peak close to the \thetap mass, from inclusive 
production.  One of these collected data from neutrino 
experiments (ITEP\cite{itep}) and two others 
used electroproduction (HERMES\cite{hermes} and 
ZEUS\cite{zeus}).  Of course, the $K_s^0$ 
is a mixture of both strangeness $+1$ and $-1$, so the 
invariant mass spectra will include both $\Sigma^{*+}$ and 
possible \thetap peaks.  It follows that a peak at a 
mass where no $\Sigma^{*+}$ resonance is known could be 
evidence for the \thetap or an unknown $\Sigma^{*+}$ resonance. 
It is also curious that these three measurements reported a  
\thetap mass which is about 10 MeV below that seen by the 
first experiments (barely compatible within the 
experimental uncertainties). Furthermore, most of the 
null evidence for the \thetap (see below) also 
measure the $p K_s^0$ invariant mass, but 
no peak is seen at the \thetap mass.  The inherent weakness 
in not knowing the strangeness of a particle, 
coupled with the uncertainty in the background which must 
include the overlapping $\Sigma^{*+}$ resonances, makes 
this evidence less convincing than exclusive measurements. 

\begin{figure}
\vspace{5.5 cm}
\centerline{ 
\includegraphics{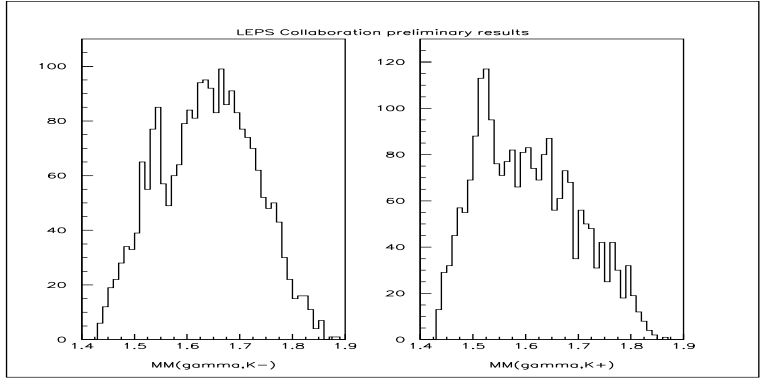}
}
\caption{ {\it Preliminary} missing mass spectra for the 
$\gamma d \to K^+ K^- X$ 
reaction measured with the LEPS detector at SPring-8.  The event 
selection requires: (1) particle ID for each kaon; (2) the missing 
mass of the $KK$ system is within 5 MeV of the nucleon mass; 
(3) a cut on the $KK$ invariant mass to remove the $\phi$-meson resonance; 
(4) the photon energy is less than 2.35 GeV.  The \thetap peak is seen at about 
1.53 GeV on the left and the $\Lambda(1520)$ peak is seen at 1.52 GeV 
on the right.
}
\end{figure}

Three experiments remain that have good evidence for the 
\thetap.  The first is from CLAS on a proton target 
\cite{kubarovsky}.  More 
details will be given separately\cite{weygand} but a few 
comments are in order.  This exclusive reaction, 
$\gamma p \to \pi^+ K^- K^+ n$ is very clean, and the 
background comes primarily from meson production reactions. 
The cuts for this analysis were not chosen arbitrary, as 
has been suggested by some critics, but are specifically 
designed to remove the dominant background along with the 
assumption that the \thetap can be produced through an 
s-channel diagram\cite{kubarovsky}. 
Furthermore, these data were examined by  
a partial wave analysis (PWA), where the amplitudes of each 
partial wave were fit over the full angular coverage of the 
CLAS detector.  Hence, the background under the \thetap 
peak (after all cuts are applied) has been fixed by the 
PWA from the full (uncut) data. The \thetap peak here 
has the highest statistical significance yet reported, 
in excess of 7 $\sigma$.  Because this is an exclusive 
measurement from the proton, there is no ambiguity in 
rescattering from other nucleons, and the strangeness of 
the final state is clearly identified.  On the other hand, 
the mass of the peak is at $1.55 \pm 0.01$ GeV, which is 
about 0.01 GeV higher than the initial \thetap measurements.

The second experiment with good evidence for the \thetap 
is the COSY-TOF result from the exclusive hadronic 
reaction $ p p \to \Sigma^+ K_s^0 p$.  Here, the 
strangeness of the $pK_s^0$ invariant mass is tagged by 
the $\Sigma^+$.  The particle identification is done 
entirely by geometric reconstruction which, for this 
near-threshold reaction, is shown to be very accurate. 
The details are given separately in these proceedings 
\cite{killian}.  Some critics have questioned whether 
this method provides good identification of the final state, 
but it can be rigorously proven that the kinematics are 
over-constrained\cite{killian}.  The result is a very 
clean final state showing a \thetap peak at a mass of 
about 1.53 GeV, which is on the low side of the \thetap 
mass measurements.

The third experiment, and perhaps the most convincing one, 
is the new data from LEPS on a deuterium target.  These 
data are shown in Fig. 1, which uses minimal cuts for 
the event selection.  There are clear peaks for both the 
$\Lambda(1520)$ and the \thetap where the same event 
sample has been used for both plots and the same Fermi motion 
correction is applied to both.  The only difference is that one 
spectrum uses the $K^+$ and the other uses the $K^-$, 
and the detector acceptance is symmetric for these 
charged kaons.  In addition, the same analysis procedures 
applied to a mixed-event test do not show any peaks at the 
location of either the \thetap or the $\Lambda(1520)$.
In the mixed-event test, the $K^+$ 
and $K^-$ are taken from different events, but the same 
analysis cuts (which ensure energy and momentum conservation) 
are applied.  Further confidence is gained by seeing that 
the peaks cannot be generated from a kinematic reflection 
of $K^+K^-$ pairs from the $\phi$-meson peak.  More details 
are given in a talk by Nakano\cite{nakano}.

\subsection{ Null Results }

Having taken a critical look at the evidence in favor of the 
\thetap we now turn to the null results.  These have 
all come from either high-energy reactions using a hadron 
beam (such as HERA-B, HyperCP, and CDF\cite{qnp}) or from 
electron-positron colliders (Belle, BaBar, BES\cite{eeexpt}). 
Because of the difficulty in detecting neutrons in these 
detectors, typically these experiments look at the $pK_s^0$ 
invariant mass, like in the HERMES and ZEUS experiments. 
Unlike the medium-energy electroproduction reactions, the 
high-energy hadron beam experiments typically have much 
higher statistics yet see no \thetap peak.  Naively, one 
might expect that if the \thetap exists, it should be 
produced in both high-energy electroproduction and 
high-energy hadron collisions, perhaps 
through fragmentation processes as the flux tube breaks 
when the struck quark exits the nucleon. This reasoning 
suggests that the \thetap does not exist. On the other hand, 
for the electron-positron collisions, it is not clear how 
an appreciable number of $\Theta$-$\bar{\Theta}$ pairs are 
produced by a mechanism where 5 quarks and 5 antiquarks 
must be produced from, say, decay of a $q\bar{q}$ meson. 

Since the hadron beam experiments pose a more serious 
challenge to the existence of the \thetap we should examine 
these experiments with some care.  In the interest of 
fairness, the same criticism directed at the HERMES and 
ZEUS experiments should also be applied to the high-energy 
hadron experiments.  Perhaps the most severe criticism is that 
the $pK_s^0$ spectra should show evidence for known $\Sigma^{*+}$ 
resonances, even if these resonances are broad, yet these 
spectra are featureless even with high statistics.  
Clearly, more effort needs to be put 
toward understanding the background in these measurements. 

The production mechanism of the \thetap (if it exists) or 
even the $\Lambda^*$ and $\Sigma^*$ resonances from 
fragmentation processes is not well known.
Hence the high-energy experiments can only put an upper 
limit on the ratio of production of, say, the pair of 
$\Sigma^*$ resonances at 1660-1670 MeV to $\Lambda(1520)$, 
or \thetap to $\Lambda(1520)$.
Non-observation of the $\Sigma^*$ resonances, which might 
be difficult to detect because of their broad width, does 
not mean that the $\Sigma^*$ does not exist.  Similar 
reasoning applies to the non-observation of a \thetap 
peak, although the limits will be more stringent because 
of its narrow width.  When examining 
the results from high-energy hadron beams, it is important 
to report not just the upper limit on \thetap but also the 
upper limits on other known hadron resonance production.
These upper limits should then be confronted quantitatively 
with calculations based on models of flux-tube fragmentation.

Finally, the facts should be clearly stated when drawing 
conclusions from both positive and null evidence.  The kinematics 
in experiments with upper limits on \thetap production are 
different from those experiments reporting positive evidence.  
In other words, the null results do not prove that the 
positive results are wrong.  There may be some interesting 
physics to be learned, assuming the experiments are correct, 
as to why exclusive measurements at medium energy show a 
potential \thetap peak whereas this signal seems to be obscured  
in high-energy inclusive measurements. 
Regardless of the explanation, the facts (and assumptions) 
should be made clear when drawing conclusions about 
the existence of the \thetap pentaquark.

\subsection{ Experimental outlook }

Two measurements expected to produce about a ten-fold 
increase in statistics, one using a deuterium target and 
the other using a hydrogen target, are currently being 
analyzed from CLAS.  However results are not expected to be 
ready until late 2004.  In addition, another experiment 
to get more statistics from the COSY-TOF detector is 
planned within the year.  High statistics will be a crucial 
test of whether the \thetap exists or not.

Analysis of the new deuterium data from CLAS is in progress. 
The missing mass spectrum for the $\gamma p \to K^+ K^- p n$ 
reaction, where only the charged particles are detected, shows 
a clean neutron peak with very litle background.  The 
photon beam had a maximum energy of 3.6 GeV, compared with 
the published data\cite{stepanyan} which was taken at two 
photon beams with 2.3 and 3.0 GeV maximum energy.  The higher 
beam energy and the lower magnetic field of the new measurement 
provide different kinematics than before\cite{stepanyan}. 
(Note that one can match the kinematics of the earlier CLAS data 
by limiting the photon energy and making angular cuts in the 
data analysis.)  Because of the importantce of ``getting it right 
the first time", the CLAS collaboration has chosen not to 
release this mass spectrum for the $nK^+$ system (where the 
\thetap peak would be expected) until the analysis results 
on the {\it full data set} are final.

Similar comments apply to the new proton data from CLAS.  
However, in this case the photon beam energy is lower than the 
published result\cite{kubarovsky}, which had the majority 
of the data taken at a maximum photon energy of 5.4 GeV. 
The new data are focussed on a measurement of the reaction 
$\gamma  p \to K^0 K^+ n$, similar to that reported by the 
SAPHIR collaboration\cite{saphir}.  Preliminary data on 
this reaction was reported at the NSTAR 2004 
conference\cite{nstarpK0} but these data have not been published 
because higher statistics were desired.  Indeed, the new data 
will provide at least a ten-fold increase in statistics in 
the desired photon energy range.

The measurement at COSY-TOF will take data with improved 
detector resolution within the next year, and could increase 
the statistics of the reported results\cite{wagner} by 
as much as a factor of five.  In addition, the new data are 
planned at a slightly higher beam energy so that the detector 
acceptance is more uniform in the region above the \thetap 
peak.  Again, this measurement could be definitive if the 
reported results are reproduced with higher statistics.  On 
the other hand, if the peak is not reproduced then it will 
be important to find an explanation of the previous 
COSY-TOF \thetap peak.

\section{ How small is the \thetap width? }

One of the most curious features of the \thetap is its 
apparent narrow width.  Because the \thetap can ``fall apart" 
(without quark pair creation) into a kaon and a nucleon, 
its width is expected to be hundreds of MeV\cite{cap,kl,jw}.  
On the other hand, the chiral soliton model predicts  
a narrow width\cite{dpp} for the \thetap based on symmetries 
of the model (along with known constants such as the pion 
decay constant).  However, the width predicted by the chiral 
solition model has significant model dependence\cite{dpp2,ellis} 
and so may not be sufficient by itself to explain a width 
as narrow as 1.0 MeV\cite{ellis}.
The quark model can also obtain a width on the order of 5-10 MeV 
if there are correlations between quarks, such as diquark 
pairs in the \thetap wavefunction\cite{kl,jw,kl2}.  If the 
\thetap exists and has a width of a few MeV or less, then this 
presents a significant challenge to the conventional quark model.

Unfortunately, only upper limits on the \thetap width are 
available from the experiments described above.  The smallest 
upper limit comes from the DIANA collaboration result, which 
shows a peak width of 9 MeV.  However, Cahn and Trilling 
\cite{cahn} have shown that the DIANA result implies an intrinsic 
resolution of the \thetap on the order of 1 MeV.  Other 
studies of the phase-shifts determined from older $KN$ 
scattering data\cite{nussinov,arndt} suggest that if the 
\thetap exists with a mass in the range of 1530 to 1550 MeV, 
then its width is likely to be about 1 MeV or less.  More 
recent preprints\cite{jeulich,gibbs} have done careful 
calculations which compare directly with the $KN$ data 
and come to a similar conclusion that the \thetap width 
is on the order of 1 MeV.  On the other hand, the $KN$ 
database is very sparse in this energy range, corresponding 
to the \thetap mass, and better data is desired\cite{barnes}. 

Two experiments have measured a \thetap width that is not just 
an upper limit.  The HERMES experiment\cite{hermes} has quoted 
a width of $13 \pm 9 \pm 3$ MeV, where the first uncertainty 
is statistical and the second one is systematic, which is larger 
than the experimental resolution of 4-6 MeV.  The \thetap width 
also depends on the underlying assumptions of the background 
shape, but in all cases the uncertainties are large enough to 
be consistent with a width of 1 MeV.  The ZEUS experiment 
\cite{zeus} measured an intrinsic width of $8 \pm 4$ MeV for 
the \thetap which is about two standard deviations away from 
a 1 MeV width.  It is worth mentioning that width quoted here 
is from a combination of $pK^0$ and $\bar{p}K^0$ data, and 
the ZEUS experiment concludes that the width is above, but 
consistent with, the experimental resolution of 2 MeV.

\subsection{ Theoretical Speculation }

If the \thetap width is really as narrow as 1 MeV, then 
both quark and soliton models will have a difficult time 
to explain such a narrow width.  
This fact alone is perhaps one of the most 
worrisome aspects to the question of \thetap existence. 
However, it is possible to explain the narrow width\cite{kl2} 
if the \thetap is a mixture of two mass eigenstates. One 
suggestion\cite{jennings} is that the diquark model of 
Jaffe and Wilczek\cite{jw} and the diquark-triquark model 
of Karliner and Lipkin\cite{kl} could mix.  In this case, 
the mixing can conspire to decouple one of the eigenstates 
from KN decay, giving it a narrow width (and the other 
eigenstate with a wide width\cite{kl2}).  Such a mechanism 
would require fine-tuning of the mixing, which is not a 
desirable feature of any model, but at least allows for a 
possible explanation of the narrow width. 

Another paradox is how the \thetap can be produced if the 
width is so narrow.  For photoproduction, the most 
straight-forward calculation of the production mechanism 
is through the t-channel, where the width is related to 
the coupling constant for t-channel kaon exchange.  If the 
width is about 1 MeV, then this diagram is suppressed. 
However, $K^*$ exchange is still possible, and this suggests 
experiments that could test this mechanism\cite{kl2}.  
Another possibility is production through a cryptoexotic 
$N^*$ resonance\cite{kl3}.  This was first suggested by 
the CLAS proton data\cite{kubarovsky}.  If this is the 
dominant production mechanism, then some experiments with 
null results could be explained\cite{kl3}.  As enticing 
as these theoretical ideas may be, the ultimate test will 
be experiment.

\subsection{ Experimental Outlook }

The best way to measure the width of the \thetap is with a 
high-resolution spectrometer, for a reaction that clearly 
identifies the strangeness in the final state.  A proposal for 
the reaction $K^+ p \to \pi^+ \Theta^+$ has been approved to 
run in 2005 at the KEK facility in Japan\cite{imai}.  The 
$K^+$ beam at momentum 1.2 GeV/c will be incident on a liquid 
hydrogen target of length 10 cm.  The $\pi^+$ particles will 
be detected in the SKS spectrometer with an expected resolution 
of about 1 MeV.  The cross section is estimated to be about 
80 $\mu$b, which would provide about 3500 counts in the \thetap 
peak. If the cross section is up to four times lower than expected, 
the \thetap should still be visible above background with about 
6 $\sigma$ statistical significance for a \thetap intrinsic width 
of 2 MeV or less. This direct measurement avoids the difficulty 
of Fermi momentum and final state interactions that are always 
present for a deuterium target.  If a clear peak is seen, this 
experiment will determine precise values for the \thetap mass 
and width.

\section{ How can the parity be measured? }

The parity of the \thetap is closely related to questions of the 
\thetap structure.  If the \thetap exists with a narrow width, 
then it likely has angular momentum $L>0$ otherwise it will just 
fall apart into a kaon and a nucleon.  Also, it is known that the 
$L=0$ partial waves of the $KN$ system are repulsive\cite{hyslop} 
and hence do not show resonance structure.  The same is true for 
the $L=1$ isovector partial wave, $P_{13}$, whereas the isoscalar 
$P_{11}$ is attractive. From this heuristic reasoning, the most 
likely parity assignment for the \thetap is positive parity with 
$J^\pi = \frac{1}{2}^+$. (Recall that the intrinsic parity of the 
$\bar{s}$ quark is negative.)

On the other hand, lattice gauge calculations suggest that the 
lowest mass eigenstate of the $udud\bar{s}$ system is just above 
the $KN$ threshold with negative parity\cite{fodor,sasaki} and 
the next mass eigenstate (with positive parity) is several hundred 
MeV higher.  However, these calculations are done in the quenched 
approximation with heavy quark masses, and further research is 
necessary to determine if chiral symmetry effects are significant 
when realistic quark masses are used\cite{fxlee}.  In fact, not 
all lattice calculations agree, as one that uses optimal domain-wall 
fermions\cite{chiu} gives positive parity for the lowest mass 
eigenvalue.  Another lattice calculations uses several different 
interpolating operators\cite{negele}, some of which may have 
a larger overlap with the \thetap wave function than others, 
and the results are mixed: for some operators a state with 
negative parity is found but not for other interpolating fields.
Although lattice calculations have made significant 
progress in past years\cite{morningstar} it is still too soon 
to get a conclusive prediction of the \thetap existence or parity 
from the lattice.

There are other indirect indications of the parity of the \thetap 
based on experiments.  The cross sections that have been reported 
for photoproduction experiments are on the order of 50-100 nb.  
Calculations based on Regge theory\cite{oh,ko,hosaka} give 
cross sections in this range for positive parity, but are a factor 
of 10 or more lower for negative parity.  In other words, if the 
\thetap is produced by a t-channel process, photoproduction 
experiments would not have the sensitivity to see a peak unless 
the parity is positive.  Of course, it is possible that the \thetap 
is produced by decay of an intermediate $N^*$ resonance in a 
s-channel process, which is not included in these calculations 
(because of the unknown $N^*$ coupling constant).  If more 
experimental information on, say, the angular distribution of 
\thetap production can be measured, the assumptions in these Regge 
model calculations can be tested.  Similarly, the COSY-TOF result 
gives a cross section of about 0.4 $\mu$b which is comparable 
with calculations for positive parity\cite{nam} but is more than 
a factor of ten too large if the \thetap has negative parity.

\subsection{ Experimental Outlook }

Clearly, a direct measurement of the \thetap parity is desired. 
For photoproduction experiments, the beam and target can be 
polarized, but even if the \thetap is produced in a state with 
known polarization, the parity of the \thetap can only be found 
with a measurement of the recoil polarization of its decay 
nucleon\cite{thomas}, which requires thousands of \thetap 
events and specialized detectors.  However, the situation is much 
more favorable for hadronic beams, where the Pauli exclusion 
principle restricts the symmetry of the wave function.  Near 
threshold, where only one or two partial waves are important, 
experiments with polarized beam and polarized target can be 
used to determine the \thetap parity.  The idea is quite simple 
\cite{thomas,meissner}.  For the reaction $pp \to \Sigma^+ \Theta^+$ 
when the beam and target spins are parallel, both isospin and 
spin components of the wave function are symmetric, allowing only  
odd angular momentum (negative parity) states. Similarly, antiparallel 
spins gives only positive parity states.  Experimentally, the 
cross section for \thetap production near threshold (where the 
$\Sigma^+$ and \thetap are in a relative s-wave or p-wave) is 
much bigger in one spin allignment than the other.  Such measurements 
are planned after 2007 at COSY.

\section{ Summary and Conclusions }

With both positive and null evidence for the \thetap from a 
variety of experiments, it is difficult to conclude whether 
the \thetap exists or not.  Furthermore, the theoretical 
difficulties to explain a possible narrow width, perhaps as 
small as 1 MeV, suggest that if the \thetap exists, it is 
very unusual indeed.  Nor is guidance from lattice 
gauge theory helpful, as some calculations show evidence for 
a negative parity resonance, one gets positive parity, and 
others do not see a resonance in either parity. So 
the job of proving or disproving the \thetap existence 
is currently an experimental task.

The question of existence of the \thetap 
will not be solved by a ``scorecard" approach.  With ten 
experiments with statistically-limited positive evidence 
and almost as many experiments with null evidence (with 
higher statistics but also uncertain backgrounds) 
the question is how to do better experiments. 
Several new experiments are on the horizon, and so the 
existence question should be answered within a year or so. 
The high-statistics experiments on hydrogen and deuterium 
using the CLAS detector have just finished, and the analysis 
of these data are proceeding rapidly.  A new experiment with 
a $K^+$ beam on a hydrogen target has been scheduled for 
2005, and calculations predict a \thetap peak with hundreds 
of counts and few-MeV resolution.  The COSY facility has 
also scheduled another run for the TOF detector, which 
could more than double their statistics.  These are just a 
few examples of the world-wide efforts to determine if the 
\thetap exists.


\section*{Acknowledgments}

I am grateful to my colleagues in the LEPS collaboration (Japan) 
and the CLAS collaboration (USA) who have contributed heavily to 
this area.  I thank Harry Lipkin, Marek Karliner, Bob Jaffe, 
Ted Barnes, Takashi Nakano and colleagues from CLAS for helpful comments.
This work was supported in part by the National Science Foundation
and the Research Center for Nuclear Physics in Osaka, Japan.

\end{document}